# Design and synthesis of aromatic molecules for probing electric-fields at the nanoscale

Sanli Faez,[a,†] Nico R. Verhart,[a] Marios Markoulides,[b] Francesco Buda,[c] André Gourdon,[b] and Michel Orrit[a]

We propose using halogenated organic dyes as nanoprobes for electric field and show their greatly enhanced Stark coefficients using density functional theory (DFT) calculations. We analyse halogenated variants of three molecules that have been of interest for cryogenic single molecule spectroscopy, perylene, terrylene, and dibenzoterrylene, with the zero-phonon optical transitions at blue, red, and near infrared. Out of all the combinations of halides and binding sites that are calculated, we have found that fluorination of the optimum binding site induces a dipole difference between ground and excited states larger than 0.5 D for all three molecules with the highest value of 0.69 D for fluoroperylene. We also report on synthesis of 3-fluoroterrylene and bulk spectroscopy of this compound in liquid and solid organic environments.

## Introduction

Single organic molecules in solid state host media have found a vast variety of applications in physics, biophysics, and chemistry [1]. Under cryogenic conditions, some organic molecules acquire a very strong zero-phonon line (ZPL) on their electronic transition at optical frequencies (about 300 THz) with a linewidth of merely 30 MHz, which is limited by spontaneous emission. The exact value of the transition energy is strongly influenced by the nano-environment. Given the high quality factor of this transition (~$10^7$), each molecule can be seen as a highly sensitive probe of its local environment [2, 3].

In 2001, one of us has suggested using organic molecules as nanoprobes to measure charge distributions in organic semiconductors [4]. In follow-up experiments, we demonstrated the optical detection of trapped charges in organic crystals [5]. Other theoretical modelings have shown the possibility of detecting charge displacements of merely 10 pm by triangulation methods [6]. More recently, we have shown that only 10 probe molecules are enough to yield a better accuracy than 3 nm for tracking of up to 6 electrons, simultaneously [7]. Puller, Lounis, and Pistolesi have investigated the possibility of displacement detection of nanomechanical oscillators based on electrostatic interactions [8] and have shown that regimes of strong back-action between the two-level system of a molecule and a clamped carbon nanotube can be realized.

These proposals, experimental advances in embedding these organic molecule in nanostructures [9], and recent progress in circuit-QED have motivated Das, Faez and Sørensen to propose and theoretically investigate a hybrid superconducting qubit (SCQ)-molecule-photon quantum system that enables direct state transfer from a stationary SCQ to a flying photon qubit [10]. State transfer is the key enabling step towards building large quantum networks, connecting individual electron-based nodes with light channels, while preserving quantum properties [11]. Their proposal extends the use of "molecular nanoprobes" from the effectively-classical detection of single electrons towards the quantum inspection of the electron's wavefunction, which is a crucial enabling technology for hybrid quantum information processing.

Direct probing of electrons with light in solid-state devices is challenging because the electronic state can be easily disrupted, considering the large energy gap between a single visible photon and the conduction electrons (few eV for photons versus few meV for electrons). Elastic scattering from organic molecules that are influenced by the electron field can act as the interfacing mechanism between electrons and photons with minimal energy dissipation: A molecule's internal states are influenced by the electron's electric field at distances of hundreds of nm. As the molecular states can be read out optically, the molecule can act as a transducer between optical and electronic energy scales.

The most important requirement for realizing these proposals is to obtain molecules with necessary Stark shifts, i.e., the spectral shift of the ZPL in response to an external electric field, without losing the crucial property of exhibiting a narrow zero-phonon electronic transition. Previous experiments have shown that this is indeed possible. For example, the dipole moment difference between electronic ground and excited states of some terrylene molecules in a para-terphenyl crystal reaches 1 Debye [12], corresponding to a linear shift of more than 3 MHz/(kV/m) [13]. This high sensitivity to electric field is induced by deformations of the, otherwise centrosymmetric, molecular orbitals by the shell of adjacent molecules in the crystal [14]. The electrostatic field of a single electron is roughly 150 kV/m at a 100 nm distance, which is enough to shift the ZPL frequency of those terrylene molecules by three times their 42 MHz linewidth [15].

Other molecular guest-host systems with equivalently high Stark factor and simultaneously narrow ZPL are terrylene in a polyethylene matrix [12] dibenzoterrylene (DBT) in an anthracene crystal [16]. However, in both of those systems, the Stark coefficients of the molecules were randomly set by their local environments. To pursue more quantitative measurement and to realize intriguing proposals such as mapping charge transport in nano-electronic devices, it is necessary to obtain several molecules with consistently large Stark coefficients.

In this article, we pursue one strategy for designing new organic dye molecules for electric field sensing based on the Stark shift of ZPL. To this end, we have considered targeted halogenation of otherwise centrosymmetric conjugated molecules and have calculated their enhanced Stark coefficients using density functional theory (DFT) modelling. We have analyzed halogenated variants of three molecules that have been investigated before by cryogenic single molecule spectroscopy: perylene, terrylene, and dibenzoterrylene, with their ZPL optical transitions at blue, red, and near infrared frequencies. Out of all the combinations of halides and binding sites that are calculated, we have found that fluorination of the optimum binding site induces a change of dipole moment larger than 0.5 D for all three molecules with the highest value of 0.69 D for fluoroperylene. We have synthesized targeted 3-fluoroterrylene molecules and characterized the solvent-dependent Stokes shift by bulk and fluorescence spectroscopy. Preliminary cryogenic measurements indicate that a Shpol'skii matrix such as tetradecane is not a suitable host for obtaining narrow and bright ZPL from this molecule.

## Experimental

The electric-field response (Stark shift) of the zero-phonon optical transition is due to the difference between the ground state and excited state dipole moments. Most of the previously investigated dyes, however, are centrosymmetric and hence present no intrinsic dipole moment. By embedding these molecules in a solid matrix, their symmetry is broken and they can obtain a permanent dipole moment. This dipole moment is strongly influenced by the slight deformation of each molecule and hence is highly inhomogeneous. The simplest way of directing the inhomogeneity toward larger values is to synthesize molecules with intrinsically broken symmetries. Synthesis of all of these molecules, however, is a cumbersome and expensive procedure. Therefore, we first used DFT calculation to choose the most promising candidates. It is also important not to fully disturb the orbital structure of the molecules and to keep some spatial overlap between the ground and excited electronic states, else the ZPL transition will be too weak or completely absent.

Previously, methylation of terrylene has been tested experimentally and the possibility of obtaining narrow ZPL had been demonstrated by hole-burning experiments [17]. The ZPL of those molecules in hexadecane matrix, however, were not completely stable and was prone to spectral jumps presumably because of the rotation freedom of the methyl group which could tunnel between two eigen-states. Based on that experience, we chose to replace one hydrogen atom with a single electronegative halogen atom for this project.

The structures of perylene, terrylene, and DBT are shown in scheme 1. For terrylene, we have considered replacement of a H-atom in the corner (position 3) with each of the four halogen atoms and found out that fluorine substitution provides the highest dipole change. Therefore, for perylene and DBT we have only considered halogenation with fluorine. For all three molecules the effect of modifying the position of Fluorine substitution has been investigated for all nonequivalent sites of each. The DFT results for absorption and emission energies and the direct transition are summarized in Tables 1 and 2. The computed absorption energy for pure terrylene is in relatively good agreement with our experimental data (see figure 1) and results reported in the literature [13]. The absorption energy for perylene is also in a very good agreement with the experimental value of ~434 nm. Note that solvent-specific interactions have not been considered in our DFT calculations. This first allowed singlet excitation in terrylene has a strong HOMO-LUMO character. As a representative example we show the HOMO and LUMO for the 3-fluoroterrylene in scheme 2. According to the calculation, the highest dipole moment difference between the ground and the excited states for fluorine substituted perylene, terrylene, and DBT are for positions 3, 7, and 2, respectively. The calculations also indicate that the main contribution to the dipole difference between the two states is the rotation of the dipole moment orientation rather than a change in its absolute value.

Based on the results of DFT calculations, the relative yield of chemical synthesis, and considering the accessible range of our laser sources for high-resolution spectroscopy, 3- fluoroterrylene (3FT) was chosen as the most proper candidate for testing our design strategy and to test the agreement between calculations and experimental outcome. We therefore, synthesized this substance and performed absorption and fluorescence emission spectroscopy on it. Bulk emission and absorption spectra of terrylene and 3FT in various liquid solvents, chosen to yield a large span of refractive indices, are depicted in Figures 1a and 1b. The peak absorption wavelength for terrylene dissolved in toluene is at 560 nm which is in relatively good agreement with the DFT result. The corresponding value for 3FT is 540 nm, exhibiting a significant blue-shift of 20 nm, as opposed to the DFT calculations that indicated a light red-shift. This discrepancy might be due to the solvent effect missing in the DFT result. The absorption spectrum of 3FT, showing a main peak followed by two characteristic vibronic components, although much broader, very much resembles the spectrum of terrylene. The emission spectrum of 3FT, on the other hand, does not resemble the mirror image of the absorption spectrum. Most of the emission seems to be in one of the vibronic components suggesting that the electron cloud in the excited state is heavily perturbed by the presence of the fluorine atom. Upon cooling the sample down to the liquid nitrogen temperatures, no sharp zero phonon line appears in the spectrum (see figure 1c). The position of the emission peak is very close to the value predicted by TDDFT/B3LYP.

### Optical Spectroscopy

Absorption and fluorescence emission spectra were recorded for 3-fluoterrylene and terrylene in a variety of solvents at ambient conditions. For the absorption measurements the sample was illuminated with white light from a xenon lamp (Mikropack, HPX-2000, High power xenon light source). Transmitted light was collected and sent to an imaging spectrometer (Horiba-Jobin Yvon iHR320, f=320 mm, 150 grooves/mm grating, LN2 cooled CCD (Symphony II detection system). To record the fluorescence emission spectra the sample was excited with 532 nm laser light (Coherent, Verdi V-5, diode pumped solid-state laser). The fluorescence emission was collected in reflection mode and sent to the spectrometer. The laser light was blocked by two 532 nm notch filters.

**Computational details**

All DFT and time-dependent DFT (TDDFT) calculations have been performed with the Gaussian 09 program [18]. We used the B3LYP exchange-correlation functional. The basis set is cc-pVDZ for all calculations except for the case with I (iodine) since this basis set is not available for this element. Instead for I we use the DGDZVP basis set, which should be of similar accuracy. TDDFT/B3LYP has been used also to optimize the geometry in the first excited state.

**Synthesis**

The synthesis of the targeted 3-fluoroterrylene **6** is shown in Scheme 4 [19]. The synthetic route started with the preparation of 2-(4-fluoronaphthalen-1-yl)-4,4,5,5-tetramethyl-1,3,2-dioxaborolane **2**, which was obtained in 58 % yield by Miyaura borylation of 1-bromo-4-fluoronaphthalene **1**, using a modified procedure to that reported by Miyaura *et al*. [20] 3-Bromoperylene **4** was obtained in 87 % yield by NBS bromination of perylene **3**, following the procedure reported by Mitchell *et al*. [21, 22] in combination with our previously reported method [23]. Subjecting 2-(4-fluoronaphthalen-1-yl)-4,4,5,5-tetramethyl-1,3,2-dioxaborolane **2** with 3-bromoperylene **4** under modified Suzuki cross-coupling conditions [23, 24] afforded the key intermediate 3-(4-fluoronaphthalen-1-yl)perylene **5** in 75% yield. The structure of **5** was confirmed by $^1$H, $^{13}$C and $^{19}$F NMR spectroscopy, and by UV and mass spectrometry.[1] Having obtained the key intermediate 3-(4-fluoronaphthalen-1-yl)perylene **5**, attention was then turned to selective cyclodehydrogenation reaction conditions. A representative cyclodehydrogenation procedure of **5** involved the use of eight equivalents of aluminium chloride in chlorobenzene and the mixture was stirred at 80 °C for 4 h. Following a standard workup the precipitate was then washed thoroughly with dichloromethane and extracted in a Soxhlet apparatus with toluene. The orange solution was decanted and the solid product was collected from the flask walls, rinsed with petroleum ether and dried in vacuum to give the targeted product 3-fluoroterrylene **6** as a purple solid in 31% yield. The structure of **6** was confirmed by $^{19}$F NMR spectroscopy, UV-vis and mass spectrometry [19].

## Discussion

Over the past 25 years, experimentalists have discovered and studied a variety of organic fluorescent molecules with excellent quantum properties [25]. Embedding these molecules in carefully chosen host crystals confers a high degree of coherence to their optical ZPL electronic transition already at moderately low temperatures (around 2 K). Other emitters have also been suggested for optical single electron sensing, most notably the nitrogen vacancy center in diamond [26]. All these emitters, however, must be embedded in extremely pure inorganic crystals which puts stringent conditions on their fabrication and makes them harder to combine with other types of electronic devices. Besides their narrow coherent transition and large Stark shift, organic molecules have three great advantages over other solid-state emitters: they are

  a. *Small*: less than 1 nm. They can be placed very close to the investigated device and can be used in concentrations as high as 1000 per cubic micrometer. Current optical super-resolution techniques with bright and stable emitters routinely provide position accuracies close to 1 nm.
  b. *Flexible*: their structure can be tailored according to the specific needs of the experiment.
  c. *Reproducible*: each molecule of a given substance is completely identical to any other molecule. Using molecules, we avoid structural inhomogeneity, which is a big hindrance in nanofabricated structures such as nanocrystals or self-assembled quantum dots.

In order to realize some intriguing proposals such as electric charge mapping or hybrid quantum interfaces between electrons and photons, it is necessary to invent new molecules with high electric field sensitivity. Given the long and expensive procedure of synthesizing new organic substances, we have first performed DFT modeling on a class of substances and have found single fluorination a plausible strategy.
Based on our calculations, the magnitude of the dipole moment stays almost constant for the ground and excited states of the lowest energy excitation in all three systems. By looking at the components of the dipole moment, we see that in all three molecules, the large change in dipole moment between ground state and excited state is due to the reorientation of the dipole. The orientation of the dipole moment is shown in scheme 3 for one of these molecules. An interesting case is DBT with a F substitution in position 2 with a change in dipole moment of about 0.5 Debye. Also for perylene with F substitution in position 3 there is a large change of about 0.7 D, which again is mainly due to a rotation of the dipole moment upon excitation.
To this end, we have synthesized only one of the simulated compounds, 3-fluoroterrylene, which was seen as the most promising candidate. Since we could not find a strong ZPL for this molecule at cryogenic temperatures, we used the method of solvent-induced shifts for measuring the Stark factor.

Previously, it has been shown that by measuring the solvent induced shift of the absorption and emission peak, it is in some cases possible to estimate the dipole moments by plotting the Stokes shift $\Delta \nu$ versus the difference between the static and optical Clausius-Mossotti factors $\Delta f(\epsilon, n^2)$, where $\epsilon$ is the static permittivity and $n$ is the refractive index at optical frequencies. The change of the Stokes shift with solvent consists of two parts: a universal part proportional to the squared difference in dipole moments of the ground- and excited states $\Delta \mu^2 = (\mu_e - \mu_g)^2$ given by:

$$\Delta\nu_{\text{universal}} = \frac{2\Delta\mu^2}{hc\varepsilon_0 a^3}\Delta f(\epsilon, n^2) = \frac{2(\mu_e - \mu_g)^2}{hc\varepsilon_0 a^3}\left[\frac{\epsilon - 1}{2\epsilon + 1} - \frac{n^2 - 1}{2n^2 + 1}\right],$$

plus a solvent specific term [27, 28, 29]. Unfortunately, although the dipole moments for fluoroterrylene are large for an aromatic molecule (see table 2), the difference Δμ is still very small compared to those of the dyes that were successfully investigated before with this method, and which all present strong charge rearrangements in the excited state. Therefore, it appears that the solvent-induced shifts are dominated by the specific interactions with the solvent, and therefore that the systematic trend with solvent dielectric properties is not large enough to appear in the plot of the shift versus the solvent polarizability. Hence we were unable to determine the dipole moment change by this method.

Before trying the other halogenated substances that we have modeled, it is possibly necessary to perform new calculations on for example other substituents like nitrile groups that might induce less distortion on the electronics orbitals of the dye molecule. More detailed molecular dynamics calculations that also include positioning of the molecule in the host crystal might be necessary to predict the best molecular design for single molecule electric field sensing. An alternative experimental route can be to use other chromophores such as a single chain of poly(2-methoxy-5-(2-ethyl-hexyloxy)-1,4-phenylene-vinylene) (MEH-PPV) that exhibit electric field response [30] and a sizable ZPL [31] or to use semiconductor nanocrystals [32] or vacancy centers in diamond [26 Dolde2011], which have comparable properties. Much work however will be necessary to obtain a natural-linewidth limited spectrum in these alternative systems, which is required for using them as quantum transducers. Given the vast variety of applications for such quantum nanoprobes and the numerous advantages of organic molecules over other types of emitters, this investigation is certainly worth the effort.

## Conclusions

In summary, we have investigated one possibility for designing new dye molecules specifically as sensitive electric field probes. To our knowledge, no previous report in single molecule spectroscopy has been devoted to this topic and all the previous reports on sensing charges and electric field with single molecule had used the matrix-induced dipole moment difference for this purpose. We have calculated the Stark factor of the ZPL for variants of three widely used molecules under cryogenic conditions in which a single hydrogen atom has been substituted by a halogen atom. Our results indicate that fluorine could be the best substituent and we have synthesized 3-fluoroterrylene for our experiment. While room temperature spectroscopy on solutions of this substance shows a large spectral shift relative to terrylene in both absorption and emissions spectra, we could not find an indication of a strong ZPL for 3FT in tetradecane. While choosing other host matrices might still be helpful in obtaining a narrow ZPL, we suspect that other design strategies might be necessary to find new candidate molecules with both strong Stark factor and strong ZPL emission.

## Acknowledgements

Nico Verhart acknowledges support from NanoNextNL (project 06D.12). Marios Markoulides acknowledges funding from the FP7 project PAMS (contract: 610441). The use of supercomputer facilities was sponsored by NWO Physical Sciences, with financial support from the Netherlands Organization for Scientific Research (NWO).

**Notes and references**

1. M. Orrit, T. Ha and V. Sandoghdar, *Chem. Soc. Rev*., 2014, **43**, 973-976.
2. F. Kulzer, T. Xia and M. Orrit, *Ang. Chem. Int. Ed.,* 2010, **49**, 854-866.
3. Y. Tian, P. Navarro and M. Orrit, *Phys. Rev. Lett.,* 2014, **113**, 135505.
4. J.-M. Caruge and M. Orrit, 2001, *Phys. Rev. B,* **64**, 205202.
5. A. a. L. Nicolet, M.A. Kol'chenko, C. Hofmann, B. Kozankiewicz, and M. Orrit, 2013 *Phys. Chem. Chem. Phys*., **15**, 4415–4421.
6. T. Plakhotnik, *ChemPhysChem*, 2006, **7**, 1699-1704.
7. S. Faez, S. J. van der Molen, and M. Orrit, *Phys. Rev. B,* 2014, **90**, 205405.
8. V. Puller, B. Lounis, and F. Pistolesi, *Phys. Rev. Lett.,* 2013 **110**, 125501.
9. S. Faez, P. Türschmann, H. R. Haakh, S. Götzinger, , and V. Sandoghdar, *Phys Rev. Lett*., 2014, **113**, 213601.
10. S. Das, S. Faez, and A. S. Sørensen, *Proceedings of SPIE,* 2014 Paper 9136-89.
11. H.J. Kimble, *Nature,* 2008, **453**, 1023-1030.
12. M. Orrit, J. Bernard, A. Zumbusch, and R. I. Personov, *Chem. Phys. Lett*. 1992, **196**, 595; Erratum: *ibid* 1992, **199**, 408.
13. F. Kulzer, R. Matzke, C. Bräuchle, C., and T. Basché, *J. Phys. Chem. A*, 1999, **103**, 2408-2411.
14. P. Bordat, M. Orrit, R. Brown, A. and Würger, Chem. Phys. 2000, **258**, 63-72.
15. S. Kummer, T. Basché, C. and Bräuchle, *Chem. Phys. Lett.,* 1994, **229**, 309-316.
16. C. Hofmann, A. Nicolet, M. A. Kol'chenko, and M. Orrit, *Chem. Phys.* 2005, **318**, 1-6.
17. A. Sigl, Chr. Scharnagl, J. Friedrich, A. Gourdon and M. Orrit, *J. Chem. Phys*. 2008, **128**, 044508.
18. M. J. Frisch, G. W. Trucks, H. B. Schlegel, G. E. Scuseria, M. A. Robb, J. R. Cheeseman, G. Scalmani, V. Barone, B. Mennucci, G. A. Petersson, H. Nakatsuji, M. Caricato, X. Li, H. P. Hratchian, A. F. Izmaylov, J. Bloino, G. Zheng, J. L. Sonnenberg, M. Hada, M. Ehara, K. Toyota, R. Fukuda, J. Hasegawa, M. Ishida, T. Nakajima, Y. Honda, O. Kitao, H. Nakai, T. Vreven, J. A. Montgomery, Jr., J. E. Peralta, F. Ogliaro, M. Bearpark, J. J. Heyd, E. Brothers, K. N. Kudin, V. N. Staroverov, R. Kobayashi, J. Normand, K. Raghavachari, A. Rendell, J. C. Burant, S. S. Iyengar, J. Tomasi, M. Cossi, N. Rega, J. M. Millam, M. Klene, J. E. Knox, J. B. Cross, V. Bakken, C. Adamo, J. Jaramillo, R. Gomperts, R. E. Stratmann, O. Yazyev, A. J. Austin, R. Cammi, C. Pomelli, J. W. Ochterski, R. L. Martin, K. Morokuma, V. G. Zakrzewski, G. A. Voth, P. Salvador, J. J. Dannenberg, S. Dapprich, A. D. Daniels, Ö. Farkas, J. B. Foresman, J. V. Ortiz, J. Cioslowski, and D. J. Fox. Gaussian 09, Revision D.01; Gaussian, Inc.: Wallingford CT, 2009.
19. M. Markoulides, C. Venturini, D. Neumeyer and A. Gourdon, *N. J. Chem.*, 2015, **in press**.
20. T. Ishiyama, M. Murata and N. Miyaura, *J. Org. Chem.*, 1995, **60**, 7508-7510.
21. R. H. Mitchell, Y.-H. Lai and R. V. Williams, *J. Org. Chem.*, 1979, **44**, 4733-4735.
22. H. Maeda, Y. Nanai, K. Mizuno, J. Chiba, S. Takeshima and M. Inouye, *J. Org. Chem.*, 2007, **72**, 8990-8993.
23. S. Nagarajan, C. Barthes, N. K. Girdhar, T. T. Dang and A. Gourdon, *Tetrahedron*, 2012, **68**, 9371-9375.
24. For a review, see: A. Suzuki, *J. Organomet. Chem.*, 1999, **576**, 147-168.
25. P. Tamarat, A. Maali, B. Lounis, and M. Orrit, M. *J. Phys. Chem. A*, 2000, **104**, 1-16.
26. F. Dolde, H. Fedder, M. W. Doherty, T. Nöbauer F. Rempp, G. Balasubramanian, T. Wolf, F. Reinhard, L.C.L. Hollenberg, F. Jelezko, F., et al. *Nat. Phys*. 2011 **7**, 459-463.
27. J. B. Birks (ed.), Photophysics of aromatic molecules, John Wiley and Sons Ltd., 1970, London, section 4.12 pages 109-119.
28. E.G. McRae, *J. Phys. Chem.* 1957, **61,** 562-572.
29. K.K. Rohatgi-Mukherjee, Fundamentals of Photochemistry, New age international publishers, 1978, New Delhi, Section 4.6 and 4.7, pages 101-106.
30. F. Schindler, J. M. Lupton, J. Müller, J. Feldmann and U. Scherf, *Nature Materials* 2006, **5**, 141-146.
31. F. A. Feist, G. Tommaseo, and Th. Basché, *Phys. Rev. Lett.* 2007, **98**, 208301.
32. S. A. Empedocles, and M. G. Bawendi, *Science* 1997, **278**, 2114-2117.

# Figures and Tables

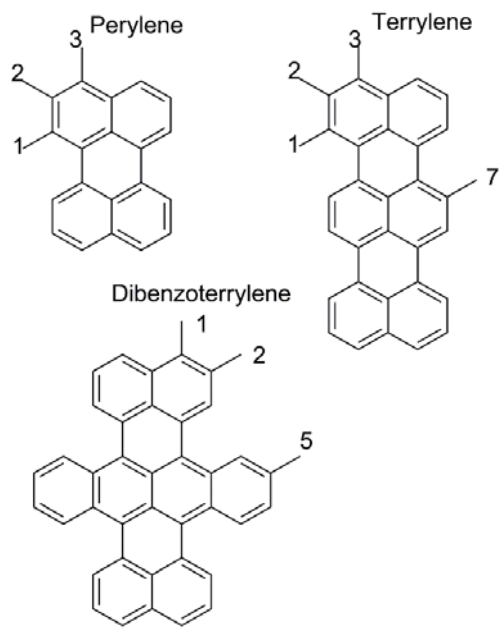

**Scheme 1**: Models systems considered in this study. The numbers refer to the positions where we have considered halogen substitution.

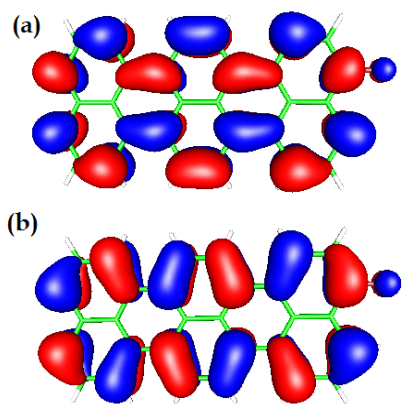

**Scheme 2**: a- LUMO and b- HOMO for 3-fluoroterrylene.

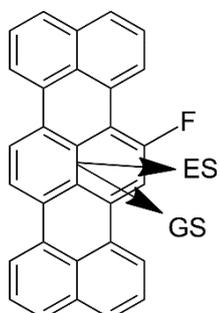

**Scheme 3**: The electric dipole moments of 7fluoroterrylene in the ground (GS) and excited (ES) states indicated that the change in the moments between the two states is mainly due to the re-orientation and not because of the change in the dipole moment magnitude.

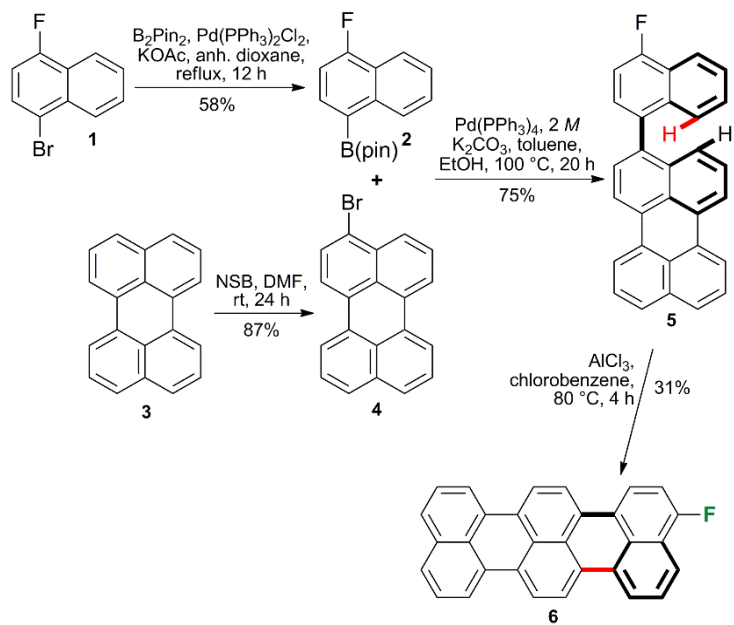

**Scheme 4**- Synthesis of 3-fluoroterrylene

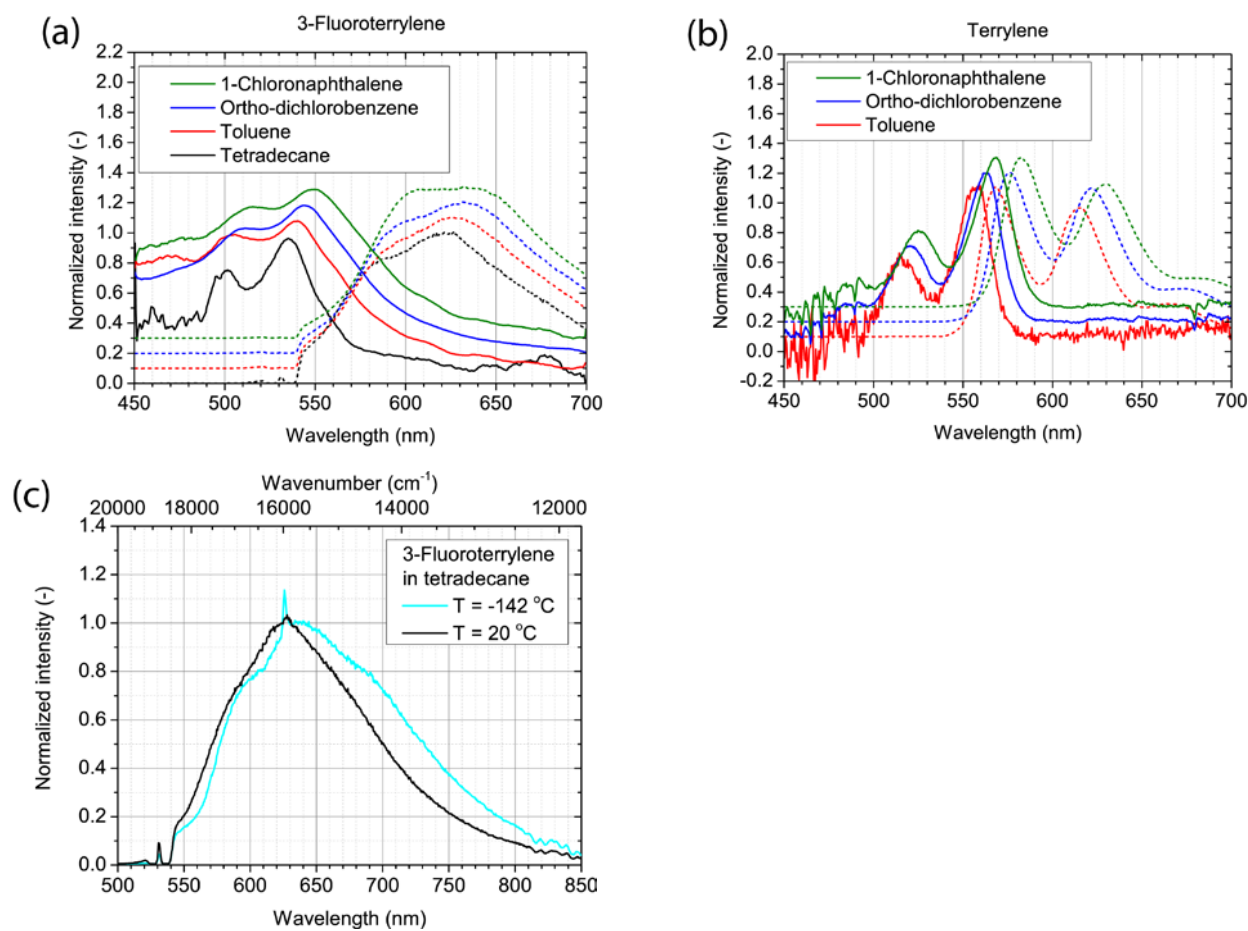

**Figure 1**: the normalized absorption- (solid curves) and emission (dashed curves) spectra of 3-fluoroterrylene (a) and terrylene (b) in several solvents. The spectra for tetradecane has zero offset, the other spectra have been cumulatively offset by 0.1 in the vertical direction for clarity. (c) emission spectrum of 3-fluoroterrylene in tetradecane at 131 K and 293 K. The sharp peak at around 625.8nm is Raman scattering of C-H stretching. There is no clear signature indicating to presence of a narrow zero phonon line.

**Table 1**: DFT/TDDFT results for dipole moments of terrylene with different halogen substitutions in position 3 (see scheme 1). The first column is the absorption energy calculated at the optimized geometry in the ground state (GS) S0 (R0); the second column is the first excitation energy computed at the optimized geometry in the first excited state (ES) S1 (R1). E-ZPL is the energy difference between the two minima in the GS and ES. The difference in dipole moment is calculated as the modulus of the difference vector between the dipole moment in the GS at R0 and the dipole moment in the ES at R1.

| Compound | Absorption | Emission | E-ZPL | GS dipole | ES dipole | Dipole change |
|---|---|---|---|---|---|---|
| | (nm) [eV] | (nm) [eV] | (eV) | (Debye) | (Debye) | (Debye) |
| terrylene | 566.0 [2.191] | 628.1 [1.974] | 2.078 | 0 | 0 | 0 |
| terrylene +F p1 | 559.6 [2.216] | 619.6 [2.001] | 2.105 | 1.10 | 1.07 | 0.20 |
| terrylene +F p2 | 566.8 [2.187] | 628.6 [1.972] | 2.076 | 1.97 | 2.14 | 0.22 |
| terrylene +F p3 | 568.1 [2.182] | 630.2 [1.967] | 2.071 | 1.38 | 1.36 | 0.36 |
| terrylene +F p7 | 559.6 [2.215] | 619.8 [2.000] | 2.104 | 1.37 | 1.35 | 0.66 |
| terrylene +Cl p2 | 567.6 [2.185] | 629.7 [1.969] | 2.073 | 2.47 | 2.70 | 0.23 |
| terrylene +Cl p3 | 573.1 [2.163] | 635.6 [1.951] | 2.053 | 2.2 | 2.21 | 0.08 |
| terrylene +Br p3 | 574.6 [2.158] | 637.2 [1.946] | 2.048 | 2.23 | 2.21 | 0.05 |
| terrylene +I p3 | 586.5 [2.114] | 653.0 [1.899] | 2.002 | 2.69 | 2.64 | 0.05 |

**Table 2**: same as table 1 for Perylene and DBT

| Compound | Absorption | Emission | E-ZPL | GS dipole | ES dipole | Dipole change |
|---|---|---|---|---|---|---|
| | (nm) [eV] | (nm) [eV] | (eV) | (Debye) | (Debye) | (Debye) |
| perylene | 436.3 [2.84] | 491.2 [2.52] | 2.68 | 0 | 0 | 0 |
| perylene +F p1 | 429.5 [2.89] | 481.9 [2.57] | 2.73 | 1.17 | 1.13 | 0.40 |
| perylene +F p2 | 437.4 [2.83] | 492.4 [2.52] | 2.67 | 1.72 | 1.94 | 0.37 |
| perylene +F p3 | 439.0 [2.82] | 494.9 [2.51] | 2.66 | 1.29 | 1.28 | 0.69 |
| DBT | 836.4 [1.48] | 975.4 [1.27] | 1.376 | 0.18 | 0.32 | 0.24 |
| DBT +F p1 | 837.9 [1.48] | 975.9 [1.27] | 1.374 | 1.25 | 1.20 | 0.13 |
| DBT +F p2 | 839.1 [1.48] | 978.7 [1.27] | 1.372 | 1.94 | 2.34 | 0.51 |
| DBT +F p5 | 837.4 [1.48] | 976.6 [1.27] | 1.374 | 1.58 | 1.52 | 0.08 |

**Table 3**: Solvent shifts for 1-fluoroterrylene and terrylene. Reported are, from left to right, the solvent, the static dielectric constant and the refractive index of the solvent, the difference between the static and optical Clausius-Mossotti factor, the peak absorption- and emission wavelengths of the fluorophore in the specific solvent and the difference (Stokes shift) and mean of the absorption and emission wavenumbers calculated from those values. The reported errors are the statistical errors obtained from the fit. The errors for the peak wavelengths of terrylene were <0.1. Note that the statistical errors are smaller than the spectral resolution of ~0.4nm of the spectrometer and the pixel resolution of ~0.5nm of the spectrometer CCD.

| 3-Fluoroterrylene | | | | | | | |
|---|---|---|---|---|---|---|---|
| Solvent | $\varepsilon$ | $n$ | $\Delta f$ | $\lambda_{abs.}$ | $\lambda_{em.}$ | $\Delta \tilde{\nu}$ | $\tilde{\nu}$ |
| | | | | (nm) | (nm) | (cm$^{-1}$) | (cm$^{-1}$) |
| 1-chloronaphthalene | 5.00[b] | 1.633[a] | 0.100 | 548.7±0.1 | 634.0±0.2 | 2451±7 | 16998±4 |
| Ortho-dichlorobenzene | 9.93[a] | 1.551[a] | 0.186 | 544.3±0.1 | 631.8±0.1 | 2544±4 | 17100±2 |
| Toluene | 2.379[a] | 1.496[a] | 0.013 | 539.5±0.1 | 626.4±0.1 | 2570±3 | 17250±2 |
| Tetradecane | 2.039[c] | 1.429[a] | ~0 | 535.2±0.1 | 621.5±0.2 | 2595±6 | 17386±3 |
| Terrylene | | | | | | | |
| Solvent | $\varepsilon$ | $n$ | $\Delta f$ | $\lambda_{abs.}$ | $\lambda_{em.}$ | $\Delta \tilde{\nu}$ | $\tilde{\nu}$ |
| | | | | (nm) | (nm) | (cm$^{-1}$) | (cm$^{-1}$) |
| 1-chloronaphthalene | 5.00[b] | 1.633[a] | 0.100 | 567.9±0.1 | 582.6±0.1 | 443±2 | 17387±1 |
| Ortho-dichlorobenzene | 9.93[a] | 1.551[a] | 0.186 | 562.6±0.1 | 576.1±0.1 | 418±1 | 17567±1 |
| Toluene | 2.379[a] | 1.496[a] | 0.013 | 557.1±0.1 | 568.3±0.1 | 354±1 | 17773±1 |

[a] Robert C. Weast (ed.), CRC Handbook of chemistry and physics, 49[th] edition, Chemical Rubber Publishing Company,

1968, Cleveland, Ohio

[b] Arthur A. Maryott and Edgar R. Smith, Table of dielectric constants of pure liquids, National Bureau of standards circular **514** (1951)

[c] Extrapolated from data for other n-alkanes (pentane, hexane, heptane, octane, etc. up to dodecane) as reported in [b].